\documentclass[twocolumn]{aastex62}

\newcommand\kms{km~s$^{-1}$}
\newcommand\msun{$M_{\odot}$}

\graphicspath{{./}{figures/}}

\received{\today}
\submitjournal{ApJ}

\shorttitle{Binary companions of Galactic supernovae}
\shortauthors{Fraser \& Boubert}

\begin{document}

\title{The quick and the dead: Finding the surviving binary companions of Galactic supernovae with Gaia}

\correspondingauthor{Morgan Fraser}
\email{morgan.fraser@ucd.ie}

\author[0000-0003-2191-1674]{Morgan Fraser}
\affil{School of Physics,
O'Brien Centre for Science North,
University College Dublin,
Belfield,
Dublin 4, Ireland.}

\author[0000-0002-0786-7307]{Douglas Boubert}
\affil{Institute of Astronomy,
University of Cambridge,
Madingley Road,
Cambridge,
CB3 0HA,
UK}



\begin{abstract}

We use {\it Gaia} Data Release 2 to search for possible surviving binary companions to three of the best studied historical Milky Way core-collapse supernovae.
Consistent with previous work, we find there to be no plausible binary companion to either the Crab or Cas A supernovae. For the first time, we present a systematic search for a former companion to the Vela supernova, and rule out essentially any surviving luminous ($>L_\odot$) companion. Based on parallax and proper motion, we identify a faint source (Star A; {\it Gaia} Source ID 5521955992667891584) which is kinematically consistent with being a former binary companion to the Vela SN progenitor. However, the inferred absolute magnitude of this source is extremely faint, raising the possibility that it may in fact be a background interloper.
In addition, we derive a new distance ($3.37^{+4.04}_{-0.97}$ kpc) to the Crab SN based on the {\it Gaia} parallax measurements, which is significantly further than the 2 kpc distance typically adopted.
Finally, we demonstrate that {\it Gaia} can be used to measure the secular decline in the luminosity of the Crab pulsar, and provide a new test of pulsar models.

\end{abstract}

\keywords{supernovae: general  --- 
parallaxes ---  pulsars: general ---  pulsars: individual (Crab, Cas A, Vela) }


\section{Introduction} \label{sec:intro}

Massive stars with a zero-age main sequence mass of more than around 8~\msun\ are thought to end their lives as a core-collapse supernova (SN). As around three quarters of massive stars are found in binary systems \citep{Sana12}, it is expected that many core-collapse supernovae will have a binary companion at the point of explosion \citep{Koch09}. Detailed calculations by \cite{Renz18} suggest that between 69 and 90\% of supernovae will have a main sequence or post-main sequence companion at the point of explosion, and that 86$^{+10}_{-22}$\% of these systems will be disrupted following the SN. We hence expect around half of all core-collapse supernovae to have an ejected stellar companion.

So far, putative binary companions have been identified for a handful of extragalactic core-collapse supernovae (e.g. SN 1993J, \citealp{Maun04, Fox14};
SN 2011dh, \citealp{Fola14};
SN 2011ig, \citealp{Ryde18};
SN 2006jc, \citealp{Maun16}). In each instance a putative companion has been detected in broadband imaging, or (in the case of SN 1993J) low S/N spectroscopy). 

In this paper, we search for surviving binary companions of massive stars which exploded as supernovae within the Milky Way. To date, such searches have focused on finding runaway stars associated with nearby SN remnants (\citealp{Guse05, Dinc15, Boub17, Kerz17, Koch18}; although see \citealp{Tetz14} for a possible companion to the pulsar PSR J0826+2637). Since {\it Gaia} Data Release 2 (DR2; \citealp{Gaia16,Gaia18}) provides exquisite 5D astrometry for the majority of point sources down to a limiting magnitude of {\it G}$\sim$20, it is now possible to trace the motion of all sources in the the vicinity of a pulsar, and search for a spatial coincidence at the point the SN would have exploded.

\section{Sample selection} \label{sect:sample}

In order to unambiguously identify the former binary companions to Galactic SNe, we require a precise parallax distance, along with measured proper motions for the associated pulsar. Furthermore, we require that the SN is sufficiently close (say, $\lesssim$10 kpc) that any companion will be bright enough for {\it Gaia} to measure its parallax, and sufficiently young (certainly less than 100 kyr) that it will still be relatively close to its birthplace on the sky.

We queried the Australia Telescope National Facility Pulsar Catalogue \citep{Manc05} for all pulsars with a measured proper motion, an age less then $100~\mathrm{kyr}$, and a distance of less than 10 kpc. This query returned 15 pulsars, of which only one (Vela) had a measured parallax.

To expand our sample, in addition to Vela we also included the Crab pulsar, along with the neutron star detected at x-ray wavelengths in Cassiopeia A (Cas A). While neither of these sources have a measured parallax from VLBI, they are sufficiently young that the area to be searched for a companion is small, rendering the constraint on distance less important.\footnote{We note that since submitting this paper to ApJ, a paper by \cite{Koch18b} was posted to the arXiv describing a search for {\it surviving} (i.e. non-disrupted) binaries in SN remnants.}
The adopted parameters for Vela, the Crab and Cas A are listed in Table \ref{tab:pulsars}.

\begin{deluxetable*}{lCCCCccCr}[tbp]
\tablecaption{Coordinates, proper motion and distances adopted for each of our neutron stars \label{tab:pulsars}}
\tablecolumns{9}
\tablewidth{0pt}
\tablehead{
\colhead{Pulsar} 
&\colhead{RA (IRCS)} 
&\colhead{Dec (IRCS)} 
& \colhead{$\mu_{\alpha} \cos{\delta}$} 
& \colhead{$\mu_{\delta}$} 
& \colhead{$\varpi$} 
& \colhead{Dist.} 
&\colhead{Age}	
& \colhead{Ref.} \\
\colhead{} & \colhead{} & \colhead{} & \colhead{(mas yr$^{-1})$} & \colhead{(mas yr$^{-1}$)} & \colhead{(mas)}& \colhead{(kpc)}  & \colhead{(yr)} &	\colhead{}
}
\startdata
Vela	& 08:35:20.611 & -45:10:34.88 & -49.68 (6)  & 29.2 (1)  & 3.5 (2) 		& -					& $\sim$11,400	& \protect\cite{Dods03}		\\
Crab	& 05:34:31.935 & +22:00:52.19 & -11.82 (22) & 2.65 (17)	& 0.27 (12) 	& -					& 961.5	& \protect\cite{Kapl08} \\
Cas A	& 23:23:27.945 & +58:48:42.45 &	-			& -	    	& -				& $3.4^{+0.3}_{-0.1}$ & $340\pm20$	&\parbox{2.8cm}{ \protect\cite{Hwan04},	\protect\cite{Reed95}}	\\
\enddata
\tablecomments{The uncertainty on the final digit(s) in $\mu_{\alpha} \cos{\delta}$, $\mu_{\delta}$ and $\varpi$ is given in parentheses. The coordinates for the Crab are taken from {\it Gaia} DR2. Ages are with respect to 2015.5, the epoch of DR2.}
\end{deluxetable*}

We note that there are in principle other young pulsars that could be included in our sample, such as PSR B1757-24 \citep{Manc85}. While PSR B1757-24 has a proper motion measurement \citep{Thos02}, it has no measured parallax from VLBI, and only weak constraints on its distance from the dispersion measure and H{\sc i} absorption. A similar situation exists for PSR B1951+32 \citep{Zeig08}. Expanding the sample of young pulsars with measured parallaxes from VLBI is hence of considerable interest.

\section{Results}

In the following section, we discuss each of our three pulsars/neutron stars in turn.

\subsection{Vela
\label{sect:vela}}

Vela is one of the closest, and youngest SN remnants. It has long been known that the remnant is associated with a pulsar \citep{Larg68}.
The age most commonly adopted for the Vela pulsar is 11.4 kyr \citep{Reic70}; which is comparable to that inferred for the SN remnant \citep{Clar76}. We note however that some authors have advocated a significantly older age of $\sim$20-30 kyr \citep{Lyne96}. Due to its proximity, we assume no foreground extinction towards Vela \citep{Fran12}.
We first search for a companion using a prior on runaway velocities from \cite{Renz18}. Following this, we consider a more general search where we do not impose a prior, but rather search for any source that intersects the past pulsar trajectory. Finally, we consider the possibility that we do not find the putative companion to the Vela pulsar as it has no parallax measurement in Gaia DR2.

\subsubsection{Searching for a companion using a prior}

\begin{figure*}[ht]
\includegraphics[width=\linewidth]{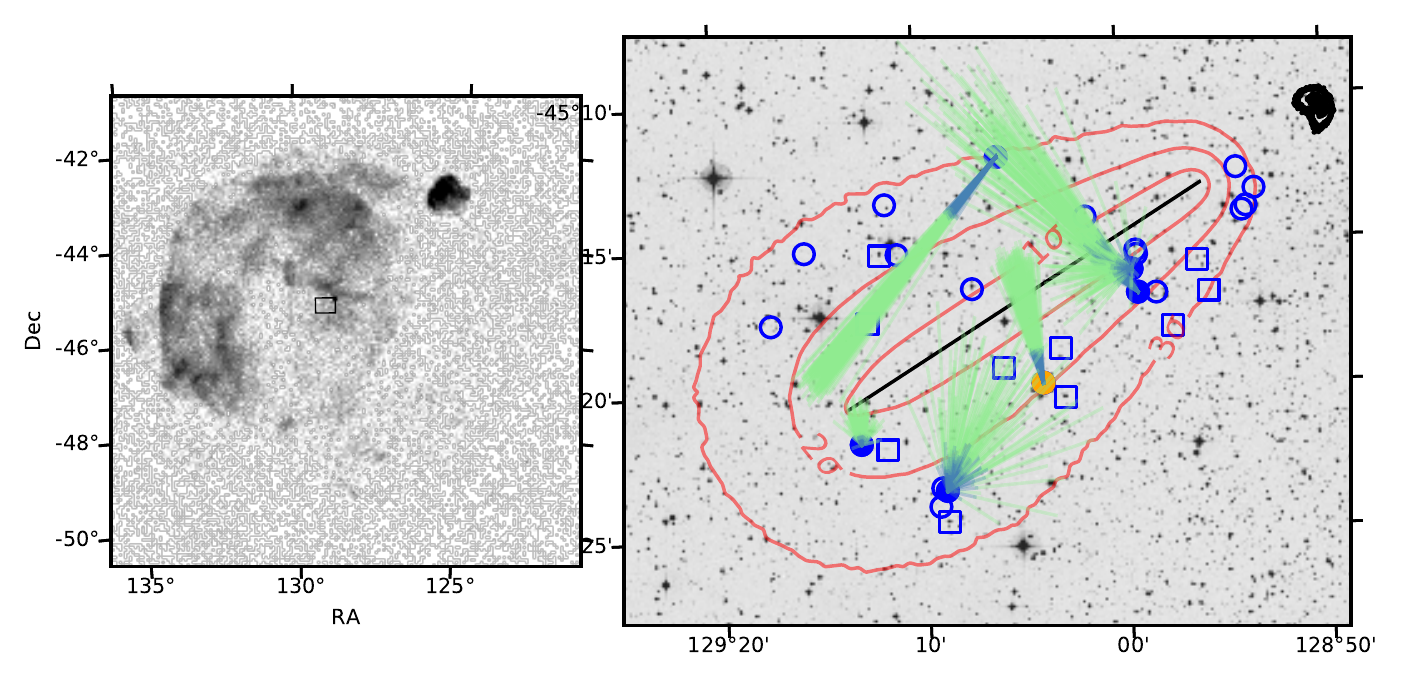}
\caption{Left panel: The Vela SNR, as seen in broadband (0.1--2.4 keV) X-rays by the ROSAT All Sky Survey \citep{Voge99}. Right panel: A zoom-in showing the region of the Vela pulsar (corresponding to the black rectangle near the center of the left panel). The background image is from the Digitized Sky Survey, while the black contours in the upper right trace the 2--8 keV x-ray emission from the pulsar and associated bow shock observed with Chandra \citep{Dura13}. The black line shows the likely location of the pulsar between 20 and 5 kyr ago.
The red contours enclose the most probable present-day locations for a former binary companion to the Vela SN. The subset of {\it Gaia} sources that are potentially at the same distance as the pulsar are indicated with open blue circles, while sources that are brighter than $G=18$ and have no measured parallax are marked with open blue squares. Finally, the six sources that potentially intersect the pulsar trajectory are marked with filled circles, with a light blue line tracing their motion over the past 5 kyr, and a green line tracing their motion over the period from 5-20 kyr. Star A is marked in orange.
\label{fig:vela_field}}
\end{figure*}

To search for possible binary companions to Vela, we first determined the likely region where a possible binary companion could be found. Using the VLBI position and proper motion of the Vela pulsar reported in \cite{Dods03}, we calculate possible coordinates $\left( \alpha, \delta \right)$ that the Vela SN exploded at, accounting for the uncertainty in $\mu_\alpha$ and $\mu_\delta$. We also account for the unknown explosion epoch, which we assume to have occurred any time between 20 and 5 kyr before the present. We tested our code by using the {\sc astropy.coordinates} package to propagate the pulsar motion forward in time from the inferred explosion location, and confirm that we recover its current position. For each explosion position, we then sample from the distribution of runway binary companion velocities calculated by \cite{Renz18}, to calculate a possible present-day position for the binary companion. After $10^7$ Monte Carlo trials, we find that any binary companion to the Vela SN would have to lie within the red contours shown in Fig. \ref{fig:vela_field}. The contours enclose 68.3, 95.5 and 99.7 \% of the Monte Carlo trials, i.e. they show the region where we would expect to find a companion at 1, 2 or 3$\sigma$ confidence.

Values of $\mu_\alpha$ and $\mu_\delta$ for the Vela pulsar were also reported by \cite{Cara01} based on {\it Hubble Space Telescope} ({\it HST}) observations. Their measurements are in significant disagreement with those of \cite{Dods03}, however, as pointed out by \citeauthor{Dods03}, the reference sources used to anchor the astrometry in the {\it HST} images are themselves moving (as they lie within the Milky Way at distances of 2-10 kpc). When  \citeauthor{Dods03} correct for this effect, the values from \citeauthor{Cara01} move closer to those found from VLBI observations. In any case, we adopt the VLBI-based positions and proper motions in our analysis, as the radio data have higher resolution than {\it HST}, and are already in a well-defined reference frame.

We queried the {\it Gaia} DR2 archive\footnote{https://gea.esac.esa.int/archive/} for all sources within a large region encompassing the probable location of any companion; and found approximately\footnote{Since we use Monte Carlo simulations to delimit where a companion may be found, the exact boundaries of this region will change slightly every time the code is run.} 5,350 sources to lie within the red contour in Fig. \ref{fig:vela_field}. To reduce this number, we used the {\it Gaia} parallaxes to discard sources that were at a distance inconsistent with the Vela pulsar.

We take the distance of the Vela pulsar to be 287$^{+19}_{-17}$ pc, from the parallax seen in VLBI observations \citep{Dods03}.
The distance is consistent with that measured from optical parallax with {\it HST} (294$^{+76}_{-50}$ pc; \citealp{Cara01}), along with that inferred from high velocity absorption lines superimposed on spectra of background sources (250$\pm$30 pc; \citealp{Cha99}).
We have no constraint on the radial velocity of the Vela pulsar, however, even if we assume that it is moving at 10$^3$ \kms\ for 20 kyr, then the distance of the pulsar would have changed by at most $\sim$20 pc. We apply distance cuts in the following that comfortably encompass this radial motion.

For each of the 5,350 sources within the red contours, and for which have a measured parallax, we calculate the distance using a Bayesian approach following the recommendations in \cite{Luri18} and \cite{Bail15}. We adopt an exponentially decreasing space density prior, and use the {\sc pyrallaxes} library to calculate posteriors for the distances. We take the mode of the posterior as our distance, and adopt the 5th and 95th percentiles as our lower and upper uncertainties. From this, we find approximately 23 sources that lie within the region where a binary companion might be found, and have a posterior on their distance that overlaps with the distance of the pulsar $\pm$3$\sigma$ (i.e. a distance range from 213 to 341 pc).

Our final selection criterion is to use the proper motions from {\it Gaia} to see whether any of these 23 sources have intersected the trajectory of the pulsar in the past. As before, we use a Monte Carlo technique and draw $10^6$ samples from the distribution of possible proper motions for both each source and the Vela pulsar, accounting for uncertainties and the known correlation between $\mu_\alpha$ and $\mu_\delta$ in {\it Gaia} data. Following the method in \cite{Boub17}, we compute the probability that each source passes within $3.843\;\mathrm{arcsec}$ (equivalent to a projected separation of $1000\;\mathrm{AU}$ at $287\;\mathrm{pc}$) of the pulsar between $5000$ and $20000$ years ago.
We find six sources where this probability is greater than $0.0001\%$, and these are listed in Table \ref{tab:vela}.

\begin{deluxetable*}{lrrrrrrrr}[tbp]
\tablecaption{{\it Gaia} DR2 sources within {\it Gaia} DR2 which are potential former binary companions. Star A is denoted with an asterisk.\label{tab:vela}}
\tablecolumns{8}
\tablewidth{0pt}
\tablehead{
\colhead{Source ID} 
&\colhead{RA (IRCS)} 
&\colhead{Dec (IRCS)} 
& \colhead{$\mu_{\alpha} \cos{\delta}$} 
& \colhead{$\mu_{\delta}$}
& \colhead{$\varpi$} 
& \colhead{D} 
& \colhead{G}	 
& \colhead{BP-RP}\\	 
\colhead{} & \colhead{} & \colhead{} & \colhead{(mas yr$^{-1})$} & \colhead{(mas yr$^{-1}$)} & \colhead{(mas)} & \colhead{(pc)} & \colhead{(mag)} & \colhead{(mag)} 
}
\startdata
5521953626147707776	  & 129.21851 & -45.36300	& -0.07 (0.92)	& -2.79 (1.20)   & 2.76 (0.52)   & $388^{+371}_{-79}$ 	& 18.87	& 1.93  \\ 
5521970359340269184	  & 129.09995 & -45.19955	& -19.54 (0.84) & 22.77 (0.87)   & 3.29 (0.49)   & $317^{+157}_{-56}$ 	& 19.73	& 3.05  \\ 
5521968229036675840	  & 128.98708 & -45.28020	& -15.43 (3.40) & -18.08 (3.62)  & 7.46 (1.50)   & $146^{+459}_{-30}$ 	& 20.73	& 1.50  \\ 
5521955992667891584*  & 129.06728 & -45.33088	& -2.75 (1.08)  & -12.89 (0.86)  & 2.82 (0.54)   & $380^{+394}_{-78}$ 	& 20.08	& 2.21  \\ 
5521968332114008192	  & 128.99177 & -45.26671	& -4.42 (2.39)  & 0.25 (2.57)    & 4.61 (1.06)   & $243^{+2614}_{-51}$ 	& 20.83	& 1.68  \\ 
5521953351269066112	  & 129.14914 & -45.39141	& 4.99 (3.89)   & -9.54 (4.39)   & 9.06 (2.51)   & $134^{+6858}_{-0}$ 	& 20.89	& 1.37  \\
\enddata
\end{deluxetable*}

Of the six candidate companions, four have a less than 1\% probability of having intersected the trajectory of the pulsar. Examination of Fig. \ref{fig:vela_field} reveals that these sources have significant fractional uncertainties on their proper motion. Furthermore, two of these sources have a 95\% upper bound to their distance of several kpc, making it quite likely that these are simply faint background sources. We hence regard these as unlikely to be associated with the Vela SN. One candidate however, \textit{Gaia}~DR2~5521955992667891584 (henceforth Star A), appears more promising with 6.8\% of the samples meeting our strict criteria. Based on its measured proper motion,  Star A was approximately coincident with the Vela pulsar $9260\pm150\;\mathrm{yr}$ ago, which is similar to when the SN was believed to have exploded. 
Furthermore, the distance towards Star A ($380^{+394}_{-78}$ pc) is in fair agreement with that inferred for the pulsar ($287^{+19}_{-17}$ pc).
However, if we take 300 pc as the distance towards Star A (ie. consistent with both the VLBI parallax to the pulsar, and the {\it Gaia} DR2 parallax to the star), then we derive a distance modulus $m-M=7.39$. This in turn implies that the absolute magnitude of Star A is $M_G=12.69$. There is one other possible candidate, \textit{Gaia}~DR2~5521953626147707776, where 4.8\% of its samples meet our criteria and it would have coincided with the pulsar $14430\pm230\;\mathrm{yr}$ ago. However, Star A is the much more likely companion of these two because the median separation at closest approach across all its samples is only $16''$, compared to $28''$ for \textit{Gaia}~DR2~5521953626147707776. 

\begin{figure}[ht]
\includegraphics[width=\linewidth, trim={0cm 0cm 1cm 1cm},clip]{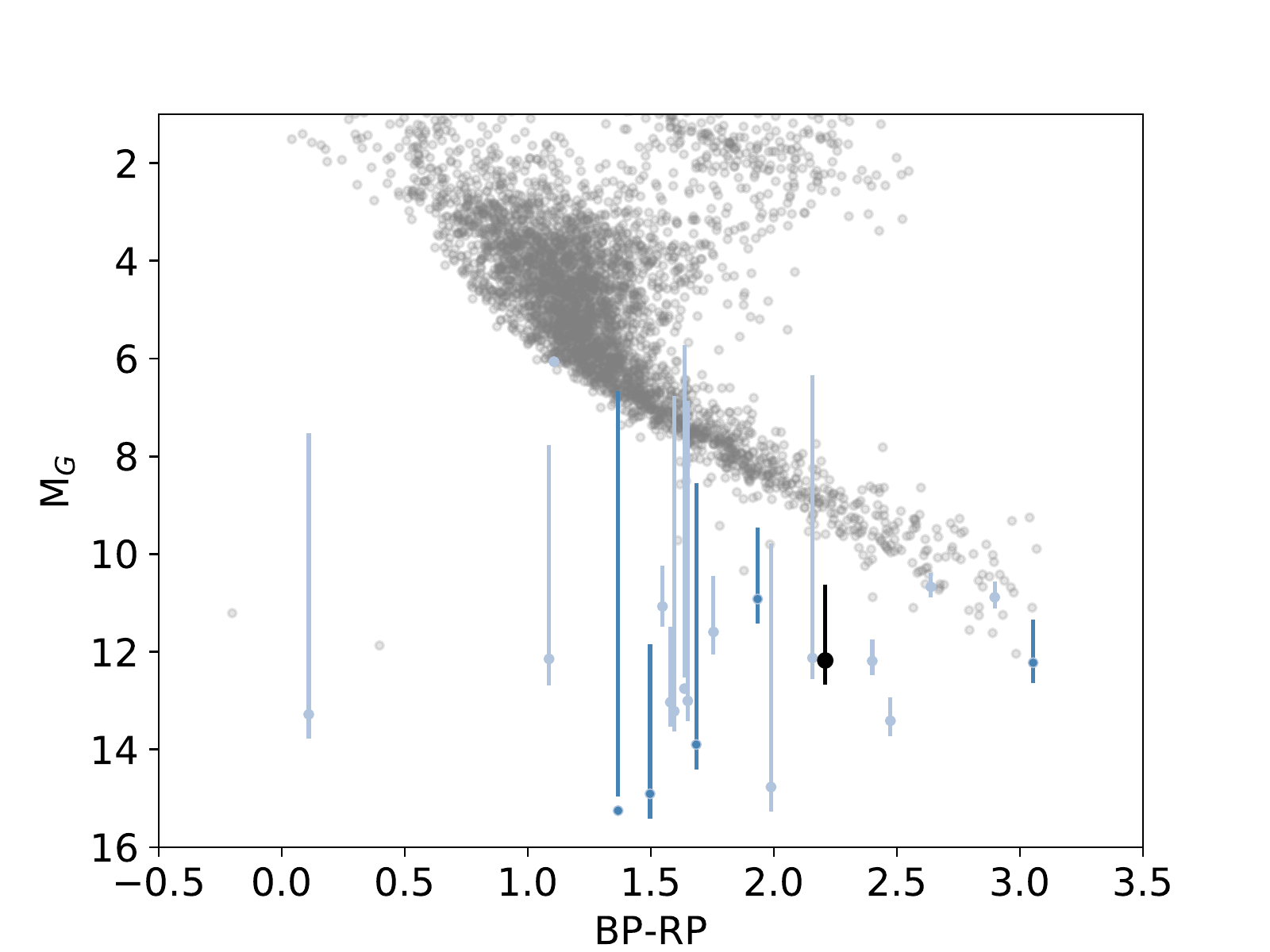}
\caption{The Hertzprung Russell Diagram for sources potentially at the distance of the Vela pulsar, and within the red contour in Fig. \ref{fig:vela_field}. Two sources are too faint to have a measured $BP-RP$ color, and these have been randomly assigned a color in the range $0<BP-RP<3$. The six sources with past trajectories intersecting that of the Vela pulsar listed in Table \ref{tab:vela} are shown in dark blue, Star A is marked in black. 
The locus of stars in {\it Gaia} DR2 are shown as light grey points in the background, these comprise of $\sim$4,000 sources selected using the criteria in \cite{Gaia18b}, and which are located within 0.5 deg of Vela on the sky.
\label{fig:vela_hrd}}
\end{figure}

A Hertzsprung-Russell Diagram is shown for the sources that are potentially at the same distance as Vela in Fig. \ref{fig:vela_hrd}. Most of these are relatively red ($BP-RP>1.5$\footnote{Gaia takes low-resolution spectra for nearly all the sources it detects with its on-board {\it Blue Photometer} and {\it Red Photometer} spectrographs. The {\it BP-RP} color is derived by integrating over the bandpass of each of these spectra, and is approximately equivalent to a $V-I$ color.}), and lie below the main sequence. However, as these have asymmetric error bars due to their uncertain distance, it is quite likely that these are considerably further than the mode of the posterior would suggest, and are hence intrinsically more luminous. In addition to these, there is one blue source ($BP-RP<1.0$) which is consistent with being a white dwarf (a handful of known white dwarfs within {\it Gaia} DR2 can be seen as the grey points in the lower left corner of Fig. \ref{fig:vela_hrd}). Star A is visible on this plot as the black point, and if we take it to lie at the upper end of its possible range of absolute magnitudes, then it is approximately consistent with being a faint and cool dwarf on the main sequence. However, in this case Star A would be too distant to be associated with the Vela pulsar.
To determine the nature of Star A, and hopefully confirm or rule out the source as a potential companion to the Vela SN, requires spectroscopy. In the absence of this, we searched for archival imaging of the field that could be used to constrain the spectral energy distribution of the star. No data were found in the {\it HST} archive, and while ground based imaging in {\it riH$\alpha$} filters was available from the VPHAS+ survey \citep{Drew14}, this did not provide additional information beyond the {\it Gaia} $BP-RP$ colors.

\subsubsection{Searching for a companion with no prior}

We also tested the effects of removing the prior on the runaway velocity (from \citealp{Renz18}), and simply performing a search for sources that were at the same distance as Vela, and intersected the past trajectory of the pulsar. For this test, we queried the {\it Gaia} archive for all sources within a 0.5\degr\ radius of (129.07, -45.33), which yielded $\sim$78,500 results. 0.5\degr\ is approximately the angular distance of a star traveling at 200 \kms\ for 11.4 kyr. We then checked whether any of these source had intersected the pulsar trajectory at any time between 5 and 35 kyr ago. This is a wider range of possible ages than what we had previously assumed for the pulsar (5-20 kyr), in order to allow for the possibility we have significantly underestimated the SN age \citep{Lyne96}. From this search, we found 13 sources within a 0.5\degr\ radius that intersected the pulsar trajectory. As expected, among these 13 sources were all of the candidates listed in Table \ref{tab:vela}. Aside from Star A, all of these sources were faint, and have a very low (1-2 \%) probability of having intersected the pulsar.

\begin{figure}[ht]
\includegraphics[width=\linewidth]{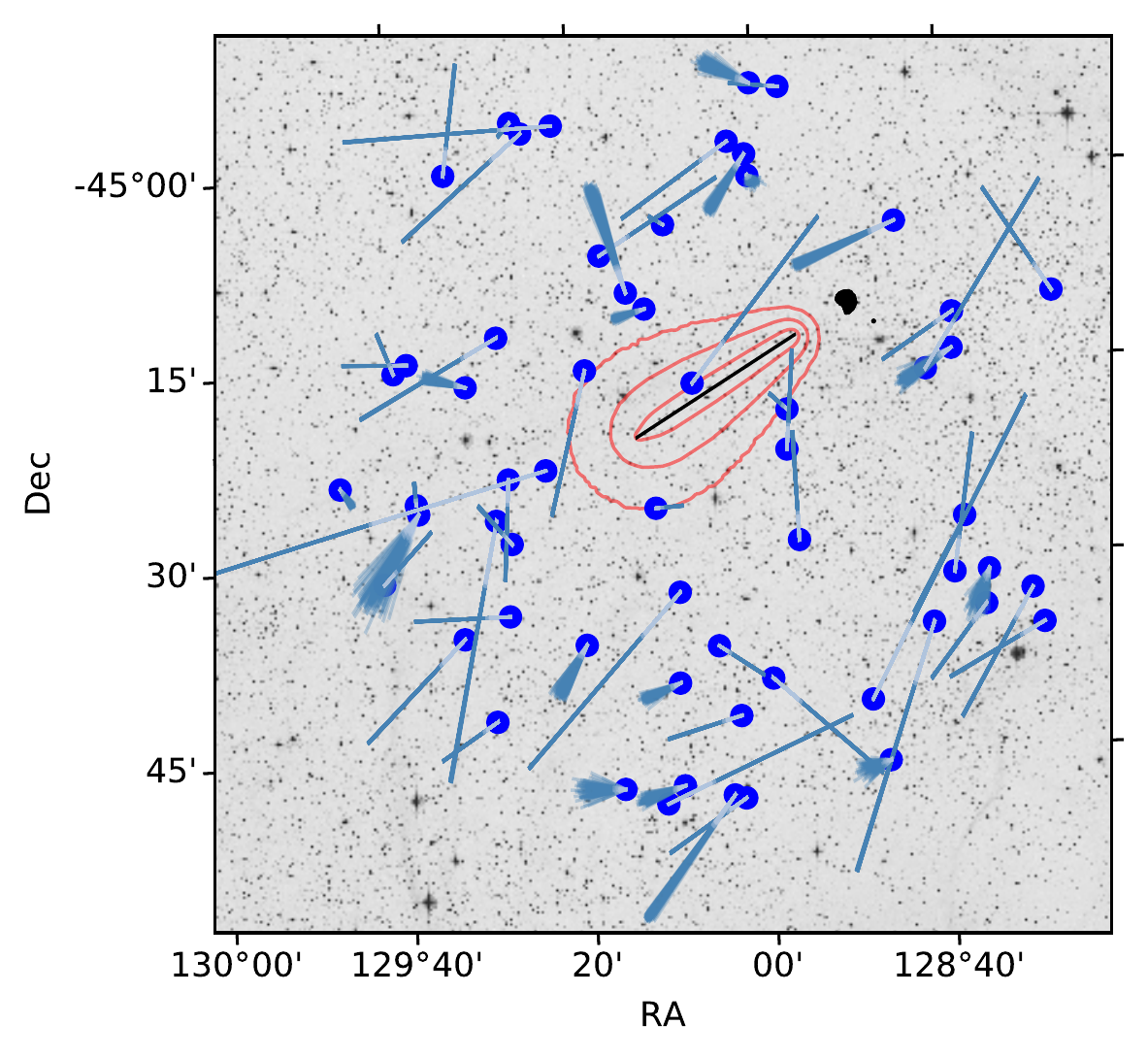}
\caption{Sources brighter than G=16, along with their inferred trajectories between 20 and 5 kyr ago (shown with blue lines). The red contours show the expected region where a companion might be found, black contours show the Chandra x-ray emission from the pulsar (as in Fig. \ref{fig:vela_field}.)
\label{fig:vela_bright}}
\end{figure}

From Fig. \ref{fig:vela_hrd}, it is evident that there are no clear over-luminous sources present in the field, as might be expected for a putative companion that had been shock heated by the impact of ejecta from the Vela SN.
As an additional test, we took all bright sources ($G<16$) that had a consistent distance, and plotted their past trajectories in Fig. \ref{fig:vela_bright}. None of these bright sources intersect the past trajectory of the pulsar, ruling them out as potential surviving companion to the Vela SN.

\subsubsection{Searching sources with no reported parallax}

Finally, we must consider whether any sources lie within the possible region where a companion could be found, but have no reported parallax in {\it Gaia} DR2. We find around 650 such sources in {\it Gaia} DR2, and most of these have magnitudes fainter than $G=20$. It is most likely that these are simply faint, distant sources for which {\it Gaia} was unable to measure a parallax. Whether a brighter source could have been missed by {\it Gaia} is somewhat harder to estimate. Around 1\% of sources brighter than mag $G=17$ in the field are missing parallax measurements in {\it Gaia} DR2. It is unclear what the reason for this is, but there are several possibilities, including marginally resolved visual binaries (where the astrometric solution may have failed to converge). It is unlikely we are missing any sources due to high proper motions - a star traveling at 200 \kms\ will have a proper motion of 0.16\arcsec yr$^{-1}$ at the distance of Vela. We hence regard it as unlikely that a companion was simply missed by {\it Gaia}.
We again note that while we have not considered the extinction towards Vela, its proximity means that $E(B-V)$ will be low \citep{Fran12}.

From the preceding analysis, we can rule out a luminous massive star as a binary companion to the Vela SN. Star A is a viable candidate due to its astrometry, however, it appears to be significantly fainter than most main sequence stars. While we must await follow-up spectroscopy of Star A, in either case any companion to the Vela SN at the point of explosion must have either been a white dwarf, neutron star or black hole, or a low mass main sequence star fainter than $L\sim 10^{-3}$ $L_\odot$.

\subsection{The Crab}
\label{sect:crab}

The supernova that gave rise to the Crab was famously observed in both China and Japan in CE 1054, and recorded as a ``guest star'' in numerous historical texts (\citealp{Duyv42}; see also the comprehensive review by \citealp{Clar77}). The detection of a young pulsar in the center of the Crab nebula \citep{Come69} indicates that this was a core-collapse SN, while the H-rich ejecta seen in the nebula points towards a Type II SN \citep[see][for a recent review of the Crab]{Hest08}.

To search for possible companions to the Crab SN progenitor, we use the same technique as for Vela. We adopt  the position and proper motion for the Crab pulsar from the {\it Gaia} DR2 catalog, and list these in Table \ref{tab:pulsars}. The Crab SN exploded in CE 1054, and we hence adopt an age for the pulsar of 961.5 years (until CE 2015.5, which is the reference epoch of {\it Gaia} DR2). The foreground extinction is likely around $A_V\sim1$ (the line of sight extinction towards the Crab at 2 kpc implies $E(B-V)=0.34^{+0.02}_{-0.02}$; \citealp{Gree18}).

\subsubsection{The distance to the Crab}

Despite being one of the best observed objects in the sky, the distance to the Crab is surprisingly poorly constrained. \cite{Tri73} reviewed the various lines of evidence and found an unweighted average distance of 1.93 kpc. Since then, and in the absence of a parallax measurement from either optical or radio \cite[see the Appendix in][for a discussion of the various observational challenges]{Kapl08}, most authors have adopted 2 kpc as the nominal distance to the Crab.

At  $V=16.7$ mag \citep{Sand09}, the Crab pulsar is sufficiently bright in optical for {\it Gaia} to measure a parallax.
While there is a ``knot'' of marginally extended optical emission located $\sim$0.6\arcsec to the south-east of the Crab pulsar (\citeauthor{Sand09}), this is distant enough for the {\it Gaia} parallax to be unaffected, and otherwise the Crab pulsar should appear as a point source to {\it Gaia}.

We queried the {\it Gaia} archive, and found only a single source ({\it Gaia} Source ID 3403818172572314624) within a 5\arcsec radius of the nominal IRCS coordinates of the Crab pulsar. The {\it G}-band magnitude of this source ($G=16.4$) is consistent with that of the Crab pulsar \citep{Sand09}. The parallax of this source is 0.27$\pm$0.12 mas, which leads to a naive $1/\varpi$ distance estimate of 3.70$\pm$1.65~kpc.

\begin{figure}[ht]
\includegraphics[width=\linewidth, trim={0cm 0cm 1cm 0cm},clip]{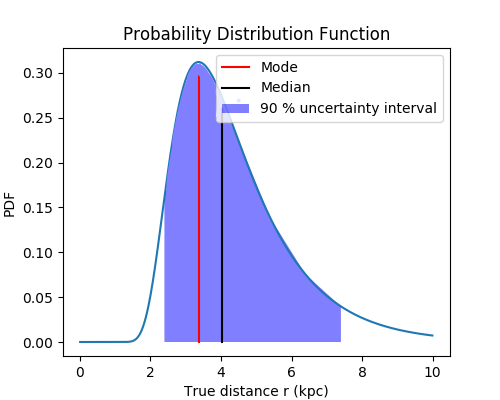}
\caption{Posterior for the distance to the Crab pulsar.
\label{fig:crab_dist}}
\end{figure}

Using the same Bayesian technique to infer distance from parallax as described in Sect. \ref{sect:vela}, we derive the distance to the Crab to be $3365^{+4038}_{-970}$ pc.
While the posterior for the distance shown in Fig. \ref{fig:crab_dist} is quite broad, we can exclude with 95\% confidence a distance to the Crab of less than 2.4 kpc.
We note that the exponentially decreasing distance prior which we employ is tuned for the ``normal'' stellar population, not pulsars. However, as the absolute magnitude of the Crab pulsar is by chance comparable to that of a typical low mass dwarf, the prior should be equally valid in this case.
We discuss the implications of this longer distance to the Crab further in Sect. \ref{sect:crab_dist}.

We also checked our distance against that reported by \cite{Bail18}, who have published a catalog of distances to all sources with measured parallax in {\it Gaia} DR2. The distance to the Crab pulsar from \citeauthor{Bail18} is 2962$_{-763}^{+1291}$ pc, where the value quoted comes from the mode of the posterior probability density, and the uncertainties encompass the $\pm 1\sigma$ range about the most probable distance. We note that the only difference between these two distances is that \citeauthor{Bail18} used a scale-length for their exponentially decreasing prior that varied with location on the sky, based on a model of the Galaxy.

\subsubsection{Searching for a companion to the Crab}

To search for a companion to the Crab, we use a similar methodology as in Sect. \ref{sect:vela}. Our starting measurements are the position of the Crab pulsar, its proper motion and parallax-derived distance from {\it Gaia} DR2. 
\cite{Kapl08} also measured proper motions for the Crab pulsar from {\it HST} images. While their value of 
$\mu_\alpha$ ($-12.0\pm0.4$ mas yr$^{-1}$) is consistent with that from {\it Gaia} ($-11.8\pm0.2$ mas yr$^{-1}$), they find 
$\mu_\delta$ to be $4.1\pm0.4$, which is significantly higher than the {\it Gaia} value of $2.6\pm0.2$ mas yr$^{-1}$.
We assume the {\it Gaia} values as more reliable in this instance as they come from a dedicated astrometric survey, whereas the {\it HST} data is taken with a number of different instruments, and each of which has a small field of view, and are consequently difficult to calibrate onto a fixed astrometric reference frame.

\begin{figure}
\includegraphics[width=\linewidth]{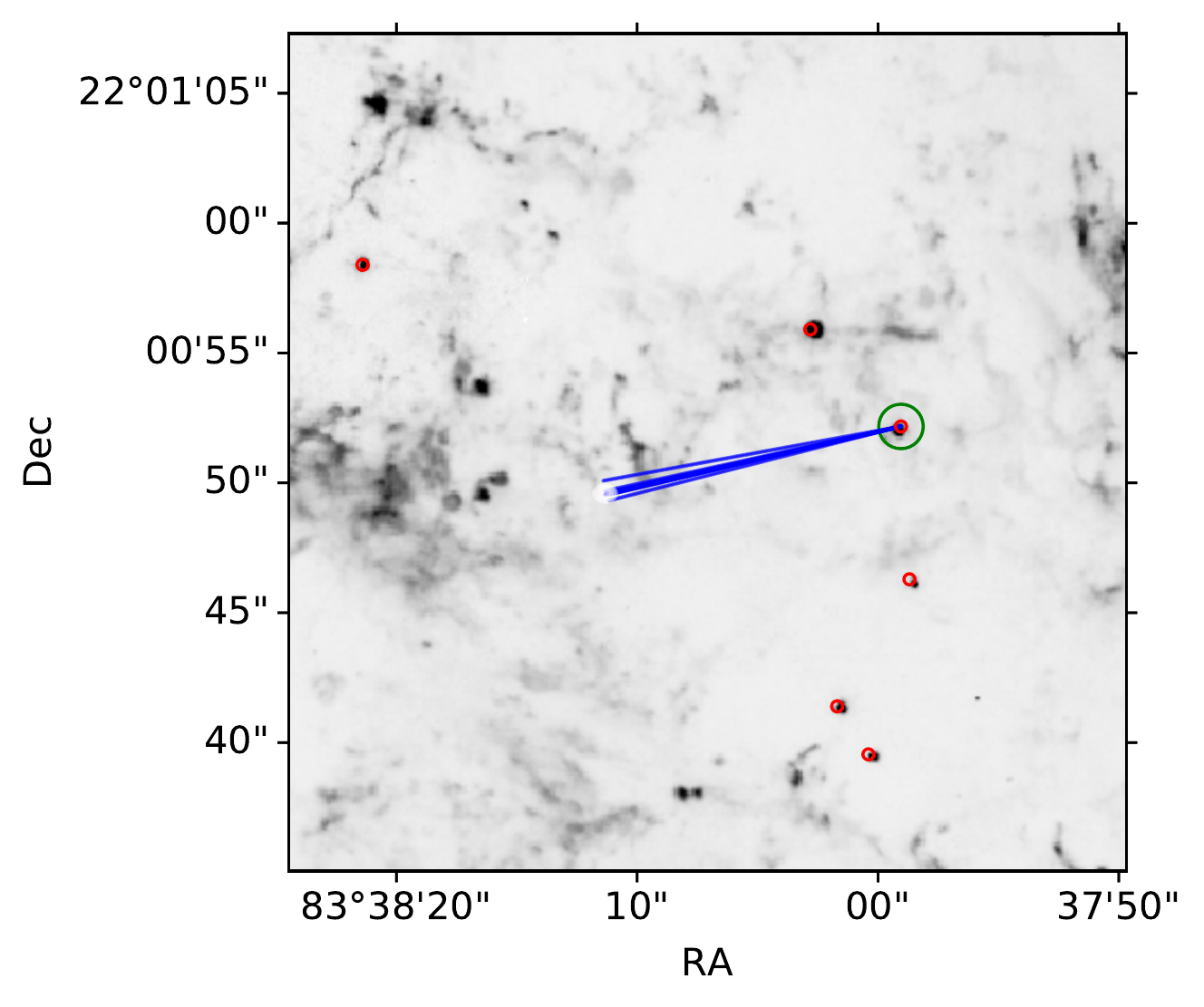}
\caption{HST WFPC2 F656N image of the center of the Crab. The pulsar is marked with a green circle, while red circles are sources in the {\it Gaia} catalog. The blue lines show a random sampling of past trajectories for the pulsar; as the pulsar has an extremely well determined age and proper motion, the blue lines all originate within a small region of radius $\lesssim 1$\arcsec.
\label{fig:crab_field}}
\end{figure}

We propagate the proper motion of the pulsar back to infer its position in CE 1054. From this point, we track how far a companion could have traveled, assuming the \cite{Renz18} velocity distribution. The results of this are shown in Fig. \ref{fig:crab_field}.

There is no source in {\it Gaia} DR2 that lies within the region where we would expect to find a companion.
Indeed, there are no sources within a radius of 10\arcsec  of our inferred explosion centre. If we adopt the traditional distance of 2 kpc to the Crab, then a star would require a velocity greater than $\sim$100 \kms\ in the plane of the sky to move beyond this. If our distance is greater than this, as the parallax to the Crab suggests, then the velocity that a companion must have had to have moved more than 10\arcsec\ from the explosion center would be even higher.

Since the posterior of the distance to the Crab is quite broad, it is not feasible to identify potential companions by searching for sources at a consistent distance. Of the 63 sources in {\it Gaia} DR2 that lie within 1.5\arcmin\ of the present day position of the Crab, over 60\% have a posterior for their distance that overlaps with that of the Crab pulsar.

To determine an upper limit to the magnitude of any surviving companion to the Crab, we take line-of-sight extinctions from \cite{Gree18}. At the canonical 2 kpc distance of the Crab, we expect an extinction of $E(B-V)=0.34^{+0.02}_{-0.02}$ mag, while at 3.37 kpc, we take $E(B-V)=0.38^{+0.03}_{-0.02}$ mag. For the line of sight towards the Crab, a distance of 3.37 kpc lies slightly beyond the range of validity for the \citeauthor{Gree18} dust map. We hence caution that the true reddening could be somewhat higher than our adopted value, if the Crab is indeed at this distance.
The limiting magnitude of {\it Gaia}, $G=20.5$, implies that for the close distance, any companion must be fainter than $M_G\sim$7.9. At 3.37 kpc, we require that $M_G$ is fainter than 6.7 mag in order to be unobservable in our data. While we cannot exclude the possibility of patchy dust that leads us to underestimate the reddening, our limits are sufficiently deep that we regard this as relatively unlikely. As for Vela, we hence conclude that the Crab SN most likely did not have a luminous ($>L_\odot$) stellar companion at the point of explosion.

The region in Fig. \ref{fig:crab_field} where we expect to find the companion to the Crab is consistent with that searched by \cite{Koch18}, who similarly found no companion.

\subsection{Cas A}
\label{sect:casa}

Cas A is one of the few Galactic SNe for which we are certain of the spectral classification (Type IIb), thanks to the detection of light echoes by \cite{Krau08}.
The absence of an unambiguous historical record of Cas A has long been noted, although \cite{Sori13} suggested that Cassini may have observed the SN on or shortly prior to 1671, while \cite{Ashw80} proposed that Flamsteed saw the SN in 1680 (although see \citealp{Step05} regarding the latter claim). The distance to Cas A, as determined from the geometric expansion of the SN remnant is $3.4^{+0.3}_{-0.1}$ kpc \citep{Reed95}. The foreground extinction towards Cas A is quite uncertain, although likely high ($A_V~5 - 15$).

The neutron star in Cas A was detected in X-rays by Chandra \cite{Tana99}, and an accurate measured position is reported by \cite{Hwan04}. While there is a measurement of the Cas A neutron star proper motion in \cite{Dela13}, it is only marginally significant, and so we do not use it in the following. As the neutron star has not been detected in radio, the prospects for measuring its proper motion in future appear remote.

We searched a 2.5\arcmin\ region centered on 23:23:27.8 +58:48:52.27 (which is the geometric center of the Cas A remnant from \cite{Reed95}; updated from B1950 to ICRS) in {\it Gaia} DR2, and found 202 sources. Of the subset of these that had measured parallaxes, 149 sources were at a distance consistent with Cas A. As these sources lie within a region approximately the same radius as the SN remnant, if they were associated with the SN then they would have had a tangential velocity that is potentially as fast as the SN ejecta. A more reasonable search radius for a companion is $\sim$10\arcsec, which is the maximum distance a companion could travel in 340 yr at 500 \kms. \cite{Kerz17} recently presented a comprehensive search for a companion to Cas A within this small region, using deep {\it HST} images, and found no candidates. A similar result was obtained by \cite{Koch18} using PanSTARRS data. The sources considered by \citeauthor{Kerz17} and \citeauthor{Koch18} are below the limiting magnitude of {\it Gaia}, and so we cannot improve upon these results.

\begin{figure}
\includegraphics[width=\linewidth]{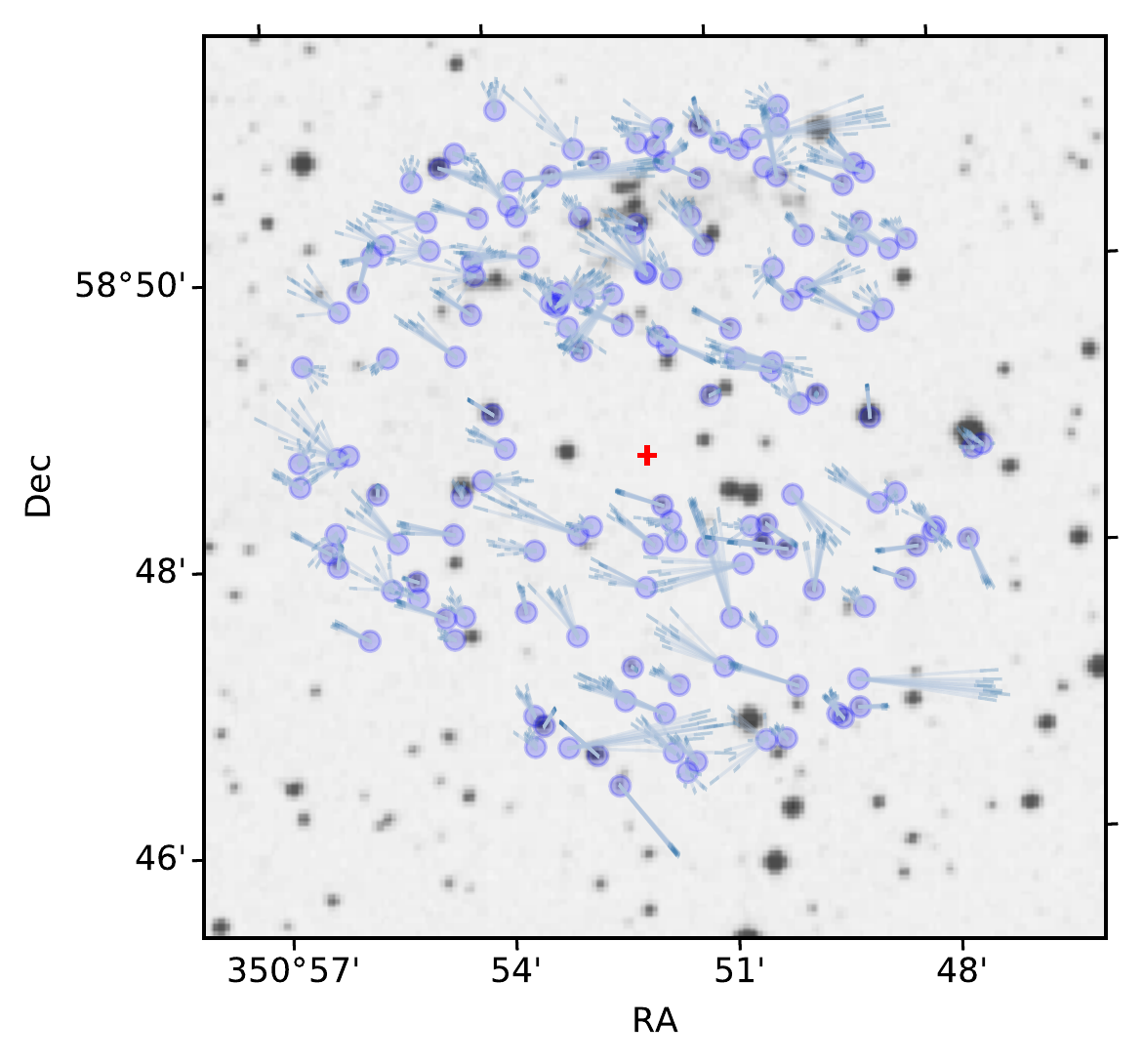}
\caption{{\it Gaia} DR2 sources with measured parallax, and that have a distance consistent with that of Cas A. The blue lines show $\times 10$ the distance the star has traveled since Cas A exploded. The red cross marks the present day location of the neutron star.
\label{fig:casa_field}}
\end{figure}

There is significant extinction towards Cas A, with estimates ranging from $A_V\sim5$ \citep[see discussion in ][]{Koch18},  or potentially as much as $A_V\sim15$ \citep{DeLo17}. For a {\it Gaia} {\it G} limiting magnitude of {\it G}=20.5, this corresponds to an absolute magnitude limit of between +2.8 and -7.2 (for $A_V=5$ and 15 respectively.
One final test we can attempt is to use {\it Gaia} DR2 to search for any bright ($G<20$) sources at the distance of Cas A that are outside the search radius of \cite{Koch18} and \cite{Kerz17}, and which have proper motions that would imply they were at the center of the remnant. In Fig. \ref{fig:casa_field} we show the proper motions of all {\it Gaia} DR2 sources in the field. From this, it is clear that there are no high proper motion stars that have trajectories implying an origin in the center of the Cas A remnant.

\section{Implications of {\it Gaia} for the Crab supernova}
\label{sect:crab_dist}

While a detailed discussion of the impact of a greater distance towards the Crab is beyond the scope of this work, we note that the absolute magnitude of the SN from the historical lightcurve will of course increase. \cite{Smit13} suggested that at peak brightness, the Crab had an absolute magnitude at peak of $M_V\sim-18$, assuming a distance of 2 kpc and an extinction of $A_V=1.6$ mag. \citeauthor{Smit13} took the apparent magnitude of the Crab SN to be $V\sim-5$, as it was recorded as being visible during the daytime.
However, it is possible that a source fainter than this could be seen in daylight. For example, on a clear day, Venus can be visible during the day, and as this is a point-like source (to the naked eye) with an apparent magnitude of $V=-4.4$, we take this as the faintest magnitude the Crab could have had that is still consistent with the historical records.
For our {\it Gaia} parallax-derived distance, and taking $E(B-V)=0.38$ from the \cite{Gree18} extinction map, we hence find an absolute magnitude at maximum for the Crab SN of $M_V = -18.2^{-1.7}_{+0.7}$, where the uncertainties reflect the uncertainties in distance and extinction towards the Crab. This absolute magnitude should strictly speaking be regarded as a lower limit, as it reflects the {\it faintest} apparent magnitude that could have been observed during the daytime, and the Crab SN may have been considerably brighter than this.
A longer distance to the Crab would also help explain its apparently low efficiency ($<<1\%$), as inferred from the ratio of the observed gamma ray luminosity to spin-down power \citep{Abdo13}. If the Crab were more distant, then its true gamma ray luminosity would be higher, implying a higher efficiency.

Finally, we note that there is one more area where {\it Gaia} can contribute to our understanding of the Crab. There is expected to be a secular decrease in the optical luminosity of the Crab \citep{Paci71} over time, as energy is lost to synchrotron radiation. A number of attempts have been made to measure this decline; \cite{Nasu96} found a decline rate at optical wavelengths of $8\pm 4$ mmag yr$^{-1}$, while \cite{Sand09} measured $2.9\pm1.6$ mmag yr$^{-1}$.
By necessity, all such measurements to date have been made with heterogeneous archival imaging taken over a timescales of decades by a number of different instruments, or rely on reported magnitudes from the literature. Differences in filter bandpasses between telescopes, along with the precise details of flux calibration, result in large and hard-to-quantify systematic uncertainties associated with these measurements \citep{Nasu96}.

In contrast, {\it Gaia} offers the opportunity to make a precise measurement of the optical decline rate of the Crab pulsar in a uniform photometric system. The {\it Gaia} mission is currently planned to last for at least seven years. Given the ecliptic latitude of the Crab, and scaling the original pre-launch estimates from 5 to 7 years, we expect that the Crab will be observed around 85 times. For each of these observations, we expect a photometric error of around 3 mmag in the broadband {\it G}-filter, given the color and magnitude ($V=16.7$; $V-I=1$) of the Crab pulsar \citep{Jord10}. The expected decline in the Crab over this timespan ($20.3\pm11.2$ mmag) should be clearly detected in the {\it Gaia} data, allowing for a new test of synchrotron emission mechanisms in pulsars.
At present, the fluxes measured for individual transits necessary for this test are provisionally scheduled for release at the end of 2022 as part of the final {\it Gaia} catalog.

\begin{figure}[ht]
\includegraphics[width=\linewidth, trim={0cm 0.0cm 1.5cm 0.5cm},clip]{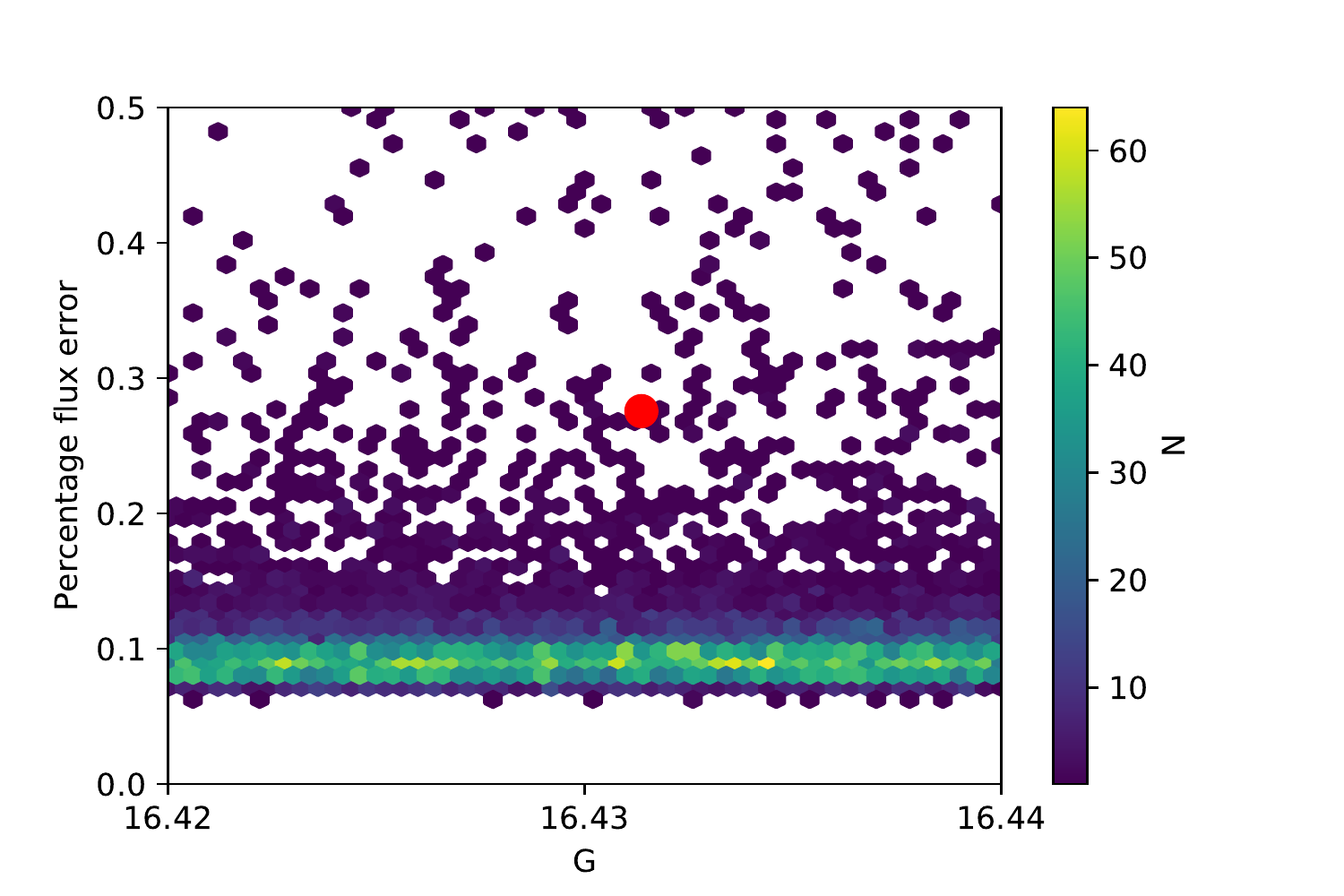}
\caption{The percentage flux error on the measurement of the Crab, compared to a density plot of 9500 other sources with the same {\it G} magnitude and number of observations ({\sc matched\_observations} and {\sc phot\_g\_n\_obs}).
\label{fig:crab_fading}}
\end{figure}

However, there is already evidence that Gaia may have detected the secular decline of the Crab in DR2. We queried the {\it Gaia} archive for all sources with the same number of observations as the Crab pulsar ({\sc matched\_observations}=15 and {\sc phot\_g\_n\_obs}=129), and where the {\it G} magnitude is within 0.01 mag of that of the Crab. In Fig. \ref{fig:crab_fading} we plot the percentage uncertainty in flux (calculated from {\sc phot\_g\_mean\_flux\_error} / {\sc phot\_g\_mean\_flux}) for each of these sources. The majority of these sources have a percentage flux error of 0.1\%, while the Crab pulsar is an outlier from the distribution. Over the 22 months of data included in {\sc Gaia} DR2, we expect the Crab to have faded by 0.6\%, and this is consistent with the Crab having a larger flux error than would be expected on the basis of its magnitude.

\section{Conclusions}

We have attempted to use {\it Gaia} DR2 to identify a former binary companion to three of the the best-studied core-collapse supernova remnants; namely Vela, Cas A and the Crab. For the latter two, we cannot improve upon extant deep literature limits. In the case of Vela, we have identified a candidate companion (Star A), however its faint absolute magnitude may be more consistent with an unrelated background source. It will be necessary to obtain spectroscopy of Star A both to confirm the spectral type of this star, and to search for possible contamination of the envelope by elements found in SN ejecta.
If Star A is shown to be unrelated, then we can exclude the presence of effectively all non-degenerate binary companions to the Vela SN.

If Star A turns out to be an unrelated background source, then there are now three historical Milky Way SNe (Vela, Cas A and the Crab), along with SN 1987A in the LMC which have no detected companion. From our initial expectation that between 69 and 90\% of stars will have a main sequence or post-main sequence companion at the point they explode, the probability of not seeing a companion for all four SNe is between 1 and 0.01\%. As noted by \cite{Koch18}, increasing the rate of stellar mergers may help to alleviate this discrepancy.

One important consideration which we have not discussed so far is the role of newly formed dust in the SN remnant, which may serve to obscure any binary companion \citep{Koch17}. While to conclusively exclude a dust-enshrouded companion would require near and mid-IR observations, we note that for the Crab and Vela the optical depth $\tau_0$ from newly formed dust is probably relatively low (as this scales with $t^{-2}$). In the case of Cas A, the remnant is perhaps more likely to be obscured due to its relative youth. Indeed, as Cas A also suffers from a high and poorly known level of foreground extinction, a search for a companion in the near infrared may be worthwhile.

Finally, we have pointed out that {\it Gaia} can contribute to studies of the Crab pulsar itself, both through a new distance estimate ($3.37^{+4.04}_{-0.97}$ kpc), and through precisely measuring the secular decline of the pulsar over time.

\acknowledgments

We thank Rob Izzard, Wolfgang Kerzendorf, Fabio Acero, David Smith and Antonio Martin-Carrillo for helpful discussions and advice. We thank the referee for their comments and suggestions.
MF is supported by a Royal Society - Science Foundation Ireland University Research Fellowship.
DB thanks the UK Science and Technologies Facilities Council for supporting his Ph.D..

This work has made use of data from the European Space Agency (ESA) mission
{\it Gaia} (\url{https://www.cosmos.esa.int/gaia}), processed by the {\it Gaia}
Data Processing and Analysis Consortium (DPAC,
\url{https://www.cosmos.esa.int/web/gaia/dpac/consortium}). Funding for the DPAC
has been provided by national institutions, in particular the institutions
participating in the {\it Gaia} Multilateral Agreement.

Based on observations made with the NASA/ESA Hubble Space Telescope, and obtained from the Hubble Legacy Archive, which is a collaboration between the Space Telescope Science Institute (STScI/NASA), the Space Telescope European Coordinating Facility (ST-ECF/ESA) and the   Canadian Astronomy Data Centre (CADC/NRC/CSA).

The Digitized Sky Surveys were produced at the Space Telescope Science Institute under U.S. Government grant NAG W-2166. The images of these surveys are based on photographic data obtained using the Oschin Schmidt Telescope on Palomar Mountain and the UK Schmidt Telescope. The plates were processed into the present compressed digital form with the permission of these institutions. 

We have made use of the ROSAT Data Archive of the Max-Planck-Institut f\"ur extraterrestrische Physik (MPE) at Garching, Germany.
We acknowledge the use of NASA's SkyView facility (http://skyview.gsfc.nasa.gov) located at NASA Goddard Space Flight Center.
     
%

\vspace{5mm}
\facilities{Gaia}


\software{astropy \citep{2013A&A...558A..33A}}

\end{document}